\documentclass{aa}

\usepackage{graphicx}

\usepackage{txfonts}

\usepackage{hyperref}
\hypersetup{
  colorlinks   = true, 
  urlcolor     = blue, 
  linkcolor    = blue, 
  citecolor    = blue  
}

\begin{document}

    \title{ZIMPOL detects scattering polarization in \ion{He}{I} D$_3$\\ during a solar flare}

   \author{Francesco Vitali \inst{1}
          \and
          Andrea Francesco Battaglia \inst{1}
          \and
          Luca Belluzzi \inst{1,2}
          \and
          Svetlana Berdyugina \inst{1}
          \and
          Renzo Ramelli \inst{1}
          \and \\
          Jiří Štěpán \inst{3}
          \and
          Gioele Janett \inst{1,2}
          \and
          Fabio Riva \inst{1,2}
          }

   \institute{
        Istituto ricerche solari Aldo e Cele Daccò (IRSOL), Faculty of Informatics, Università della Svizzera italiana (USI), 6605 Locarno, Switzerland
        \and
        Euler Institute, Università della Svizzera italiana (USI), 6900 Lugano, Switzerland
        \and Astronomical Institute of the Academy of Sciences, Ondřejov, Czech Republic\\
        \email{francesco.vitali@irsol.usi.ch}
             }

   \date{Received XX month 2025 / Accepted YY month 2025}

\abstract
   {Spectropolarimetric observations of solar flares in the \ion{He}{I} D$_3$ line at 5876~\AA~are extremely rare, and their diagnostic potential remains largely unexplored.}
   {We report the first unambiguous detection of linear polarization in \ion{He}{I} D$_3$
   during a solar flare. Using the high-precision ZIMPOL polarimeter at the IRSOL observatory in Locarno (Switzerland), we tracked the temporal evolution of the \ion{He}{I} D$_3$ Stokes profiles throughout the M7 GOES-class flare that occurred on 3 May 2023 at 10:45 UT.}
   {We analyzed the time evolution of the maximum in linear polarization and the absorption depth of the intensity profile. Both the fractional linear polarization, which peaks at $6\times10^{-4}$, and the absorption depth increase rapidly before gradually decaying, with their maxima occurring approximately 5 minutes after the peaks in GOES X-ray flux and SDO/AIA 304~\AA~emission. From the evolving \ion{He}{I} D$_3$ core position, we also derived the temporal evolution of the plasma bulk velocities.}
   {The intensity profiles exhibiting strong absorption seems to originate from the flare ribbons. The time evolution of all Stokes parameters in the 3 May 2023 event was driven by changes in ortho-helium density prompted by the different phases of the flare. Our analysis suggests that the observed \ion{He}{I} D$_3$ linear polarization is likely not dominated by the theorized impact polarization, as it exhibits neither spatial correspondence with electron precipitation sites nor temporal synchronization with the impulsive phase. Instead, the signals are consistent with scattering polarization produced by anisotropic radiation pumping.}
   {We conclude that scattering polarization signal on the order of $0.01\%$ can be produced in the \ion{He}{I} D$_3$ line during solar flares. This can provide constraints for flare models.}

   \keywords{
    Sun: solar flares --
    Sun: chromosphere --
    methods: observational --
    techniques: polarimetric --
    techniques: spectroscopic
               }

   \maketitle 

\section{Introduction}
Spectropolarimetric observations are commonly employed to diagnose the magnetic field throughout the solar atmosphere.
In solar flares, spectropolarimetry could also be used to detect nonthermal collisions \cite[]{fineschi1992electron} or the presence of anisotropic illumination \citep{vstvepan2013scattering}. 
Both of these mechanisms are sources of linear polarization, with the first producing the so-called impact polarization and the second generating resonance (or scattering) polarization.
Spectropolarimetric observations during a solar flare have been conducted in different atomic lines, such as \ion{He}{I} 10830~\AA~\citep[e.g.,][]{anan2018measurement}, \ion{Ca}{II} 8542~\AA~\citep[e.g.,][]{kuridze2018spectropolarimetric}, and $\text{H}\alpha$ 6563~\AA~\citep[e.g.,][]{YAKOVKIN20224408}. These studies have been employed to estimate the evolution of the magnetic field and the plasma temperature during solar flares.

Attempts have been made to detect impact polarization in solar flares with observations in the $\text{H}\alpha$ line. 
Although positive detections have been reported \citep[e.g.,][]{vogt1999observations}, later observations with instrumentation optimized to avoid seeing-induced cross talk did not reveal any significant linear polarization in the $\text{H}\alpha$ line during solar flares \citep{bianda2005absence}. Similarly, \citet{kawate2019infrequent} reported dozens of solar flare observations in the $\text{H}\alpha$ line, with only one showing significant linear polarization, although without spatial or temporal correlation with X-ray emissions. As was pointed out by \citet[]{bommier1986linear}, the $\text{H}\alpha$ line is sensitive to collisional depolarization, which could explain the absence of impact polarization detection during flares \citep[]{Stepan2007impactpolarization}. 

During an X1 GOES-class flare, \citet{judge2015helium} observed fractional linear polarization
in the \ion{He}{I} 10830~\AA~line reaching a few percent, but without spatiotemporal correlation with the hard X-ray (HXR) detected by the Reuven Ramaty High Energy Solar Spectroscopic Imager \citep[RHESSI;][]{lin2002rhessi}. Their analysis suggests that the observed
polarization is the result of anisotropic radiation pumping.

In this context, the polarization of the \ion{He}{I} D$_3$ line at 5876~\AA~during solar flares is largely unexplored.
This line is of particular interest because it forms in the upper chromosphere of active regions \citep[ARs;][]{mauas2005helium}. According to the standard flare model \citep[CSHKP;][]{Carmichael1964C,Sturrock1966S,Hirayama1974H,KoppPneuman1976KP}, this solar atmospheric layer is particularly relevant for solar flares because it is the region where most of the energy is released from the impact of charged particles accelerated through magnetic reconnection \citep[e.g.,][]{Fletcher2011,benz2017review}. Neutral helium consists of a singlet state (para-helium) and a triplet state (ortho-helium) between which radiation transitions are not possible, since such transitions do not allow total spin change \citep[e.g.,][]{foot2005atomic}. For this to occur, photoionization-recombination, collisional excitation, or collisional ionization-recombination are necessary physical mechanisms. This is particularly interesting in  flaring regions, because a flare can provide additional energy to populate the triplet state through ionization recombination \citep{libbrecht2021line} or nonthermal collisional ionization, as was simulated in \citet{kerr2021he}.

So far, spectropolarimetric observations in \ion{He}{I} D$_3$ during solar flares have been used for the study of the magnetic field \citep{libbrecht2019chromospheric, lozitska2024unique} and plasma velocities \citep{libbrecht2019chromospheric}. 
Since no unambiguous linear polarization signal in \ion{He}{I} D$_3$ during a solar flare has been reported prior to this study, there is no clear idea about the amplitude and shape of the signal that can possibly be observed, and the exact mechanism producing it.

In this paper, we report the first unambiguous detection of
\ion{He}{I} D$_3$ linear polarization during a solar flare. Using the high-precision Zurich Imaging Polarimeter \citep[ZIMPOL;][]{ramelli2010zimpol} at the IRSOL observatory in Locarno (Switzerland), we tracked the temporal evolution of the \ion{He}{I} D$_3$ Stokes profiles throughout the M7 GOES-class flare that occurred on 3 May 2023 at 10:45 UT.
We analyzed the time evolution of the maximum in linear polarization, of the absorption depth of the intensity profile,
and of the inferred plasma bulk velocities.
To further validate our analysis, we also present a complementary intensity observation made with the Fabry-Pérot setup during an M1 flare that
occurred on 15 November 2024.
The paper is organized as follows. Section~\ref{sec:observations} presents the instrumentation and the observations. Section \ref{datan} outlines the data analysis, while Sect.~\ref{sec:results} exposes the results. Section~\ref{sec:discussions} discusses the main findings and their implications. Finally, Sect.~\ref{sec:conclusions} provides concluding remarks.

\section{Observations}
\label{sec:observations}
\subsection{3 May 2023 M7 event with slit-spectrograph}
\label{23overview}
\begin{figure}
    \centering
    \includegraphics[width=\linewidth]{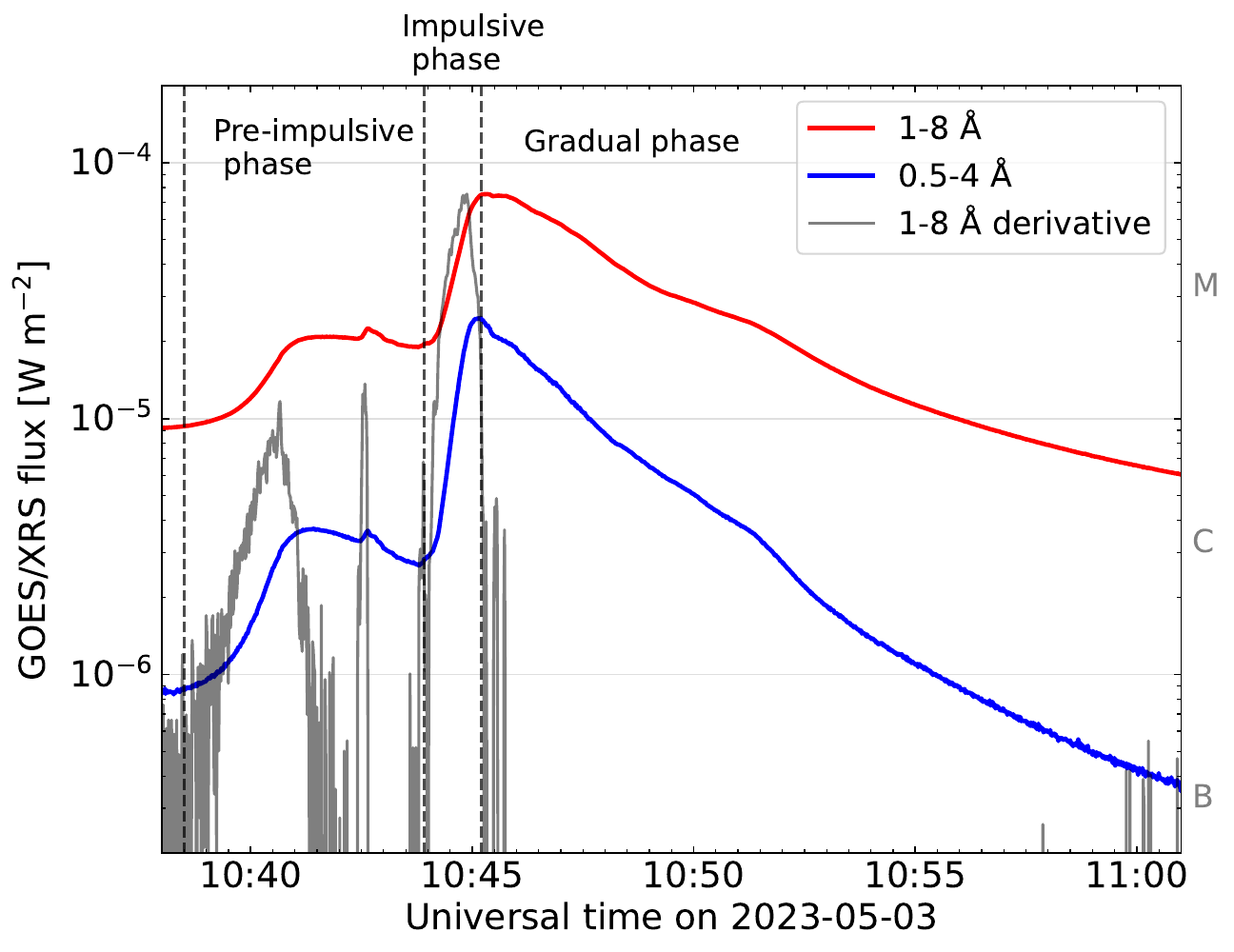}
    \caption{GOES/XRS time profiles for the
    3 May 2023 M7 event. Vertical dashed lines separate the temporal phases of the solar flare. The blue curve corresponds to the 0.5-4~\AA~GOES channel, the red curve to the 1-8~\AA~channel, and the gray curve to the derivative of the 1-8~\AA~curve.}
    \label{fig:goes-time-profiles}
\end{figure}

\begin{figure*}
    \centering
    \includegraphics[width=\linewidth]{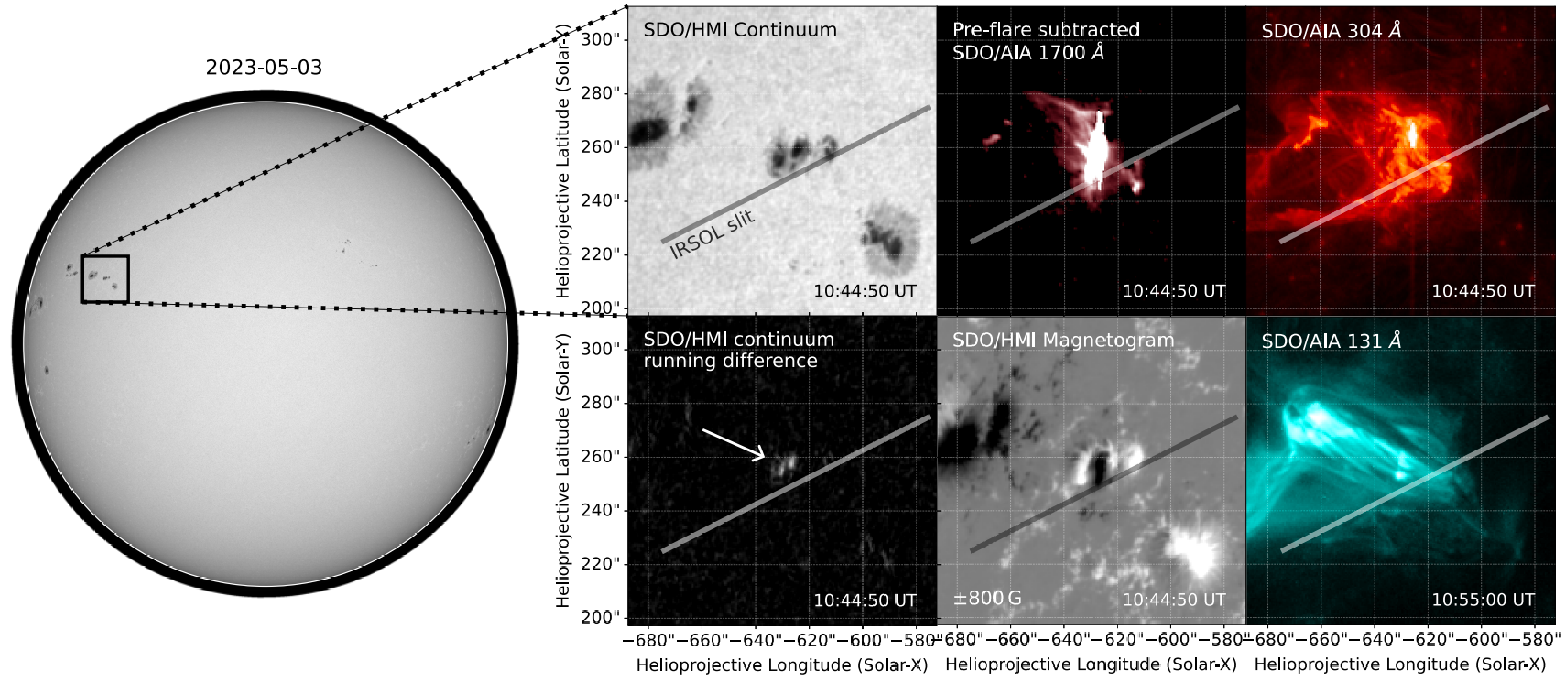} 
    \caption{Location and morphology of the 3 May 2023 M7 event. Left panel: Location on the solar disk. Right panel: Context images showing (left to right, top to bottom) SDO/HMI continuum, SDO/AIA 1700\,{\AA} (pre-flare subtracted, see Sect.~\ref{aiastix}), SDO/AIA 304\,{\AA}, SDO/HMI continuum running differences (see Eq.~\eqref{rundif}), SDO/HMI magnetogram, and SDO/AIA 131\,{\AA}. All of the images were taken at the time of the GOES X-ray flux peak, except SDO/AIA 131\,{\AA}, which was taken about 10 minutes after.}
    \label{fig:overview-figure}
\end{figure*}

Most of the present work focuses on an M7 GOES class flare observed on 3 May 2023, with the X-ray flux peak time at 10:45 UT. The time profiles of the GOES X-ray sensor (XRS) are shown in Fig.~\ref{fig:goes-time-profiles}. The flare occurred near the eastern limb of the AR 13293, as is shown in Fig.~\ref{fig:overview-figure}. The AR of the observed solar flare consists of three groups of sunspots, with the middle one being smaller in area but having stronger magnetic fields of opposite polarities. 
As is shown by Fig.~\ref{fig:overview-figure}, the flare occurred roughly across the polarity inversion line of the central group. Strong brightening from the flare is clearly observed by the Solar Dynamics Observatory \citep[SDO;][]{2012SoPh..275....3P} in the
Atmospheric Imaging Assembly \citep[AIA;][]{2012SoPh..275...17L} 1600~\AA~channel, while the 131~\AA~and 304~\AA~channels trace well the location of the flare loops.
In the Helioseismic and Magnetic Imager \citep[HMI;][]{2012SoPh..275..207S}
running differences (see Fig.~\ref{fig:overview-figure} and Eq.~\eqref{rundif}) we can see the presence of white light (WL) emission in a compact region about 10 arcsec north relative to the slit position.

The observations were carried out at the IRSOL observatory in Locarno (Switzerland). The facility hosts a Gregory-Coudé solar telescope with a  45-cm primary mirror, coupled with the Zurich Imaging Polarimeter \citep[ZIMPOL;][]{ramelli2010zimpol, gandorfer2004solar}.
This instrument is optimized to achieve high polarimetric sensitivity, which is crucial for capturing the faint polarization signals that can be found in \ion{He}{I} D$_3$ during solar flares. Thanks to its fast modulation, seeing induced errors are suppressed. Consequently, ZIMPOL achieves high polarimetric sensitivity. For this observation, we utilized the photo-elastic modulator (PEM), which operates at a modulation frequency of 42 kHz. Additionally, we employed a slow modulation device, the Telescope Calibration Unit \citep[TCU;][]{zeuner2022enhancing}, which allows for suppression of telescope polarization cross-talks as well as other systematics. To obtain data with high spectral resolution, we observed with the slit-spectrograph setup. 

We were able to measure the entire temporal evolution of the flare, from the pre-impulsive phase to the end of the gradual phase. We collected a total of 20 slit-image frames spanning 29 minutes from 10:31 to 11:00 UT. The cadence of the observations is 1.3 minutes and the camera exposure is 4 seconds for each slit-image frame. The observed spectral range goes from 5871.6~\AA~to 5900.8~\AA, with a spectral resolution of 0.046~\AA. This range includes \ion{He}{I} D$_3$ as well as the \ion{Na}{I} D$_2$ 5890~\AA~and D$_1$ 5896~\AA~lines. However, only \ion{He}{I} D$_3$ was analyzed in this work. The field of view (FOV) along the slit amounts to 200 arcsec with a spatial resolution of 2.8 arcsec. Slit-jaw images in $\text{H}\alpha$ were taken to help contextualize the observation, particularly to improve the estimation of the position of the slit on the Sun. In the ZIMPOL polarimetric reference system, the direction of positive Stokes \textit{Q} is parallel to the slit.

\subsection{15 November 2024 M1 event with Fabry-Pérot}
Although the \ion{He}{I} D$_3$ line is generally very weak in the quiet Sun, flare observations have revealed it in both absorption and emission \citep[e.g.,][]{Liu2013HeID3flare,libbrecht2019chromospheric,lozitska2024unique}. However, it remains unclear which physical mechanism is responsible for the emission and which one is responsible for the absorption.
An additional goal of this work is to investigate the origin of the absorption structure in the \ion{He}{I} D$_3$ that is formed during solar flares.

In order to compare the spatial structure traced by the distribution of the \ion{He}{I} D$_3$ absorption profiles with the flare geometry, we additionally performed flare observations with the IRSOL Fabry-Pérot system \citep[e.g.,][]{kleint2011fabryperot}, which allowed us to obtain 2D \ion{He}{I} D$_3$ intensity images. On 15 November 2024, we observed an M1 GOES class flare from AR 13893, with its peak occurring at 12:15 UT. Using the Fabry-Pérot system, we captured the rising phase of the flare, when flare-accelerated particles interact with the chromosphere and photosphere, losing energy through collisions (e.g., heating) and radiation \citep[e.g.,][]{Fletcher2011,benz2017review}. The Spectrometer/Telescope for Imaging X-rays \citep[STIX;][]{krucker2020stix} on board Solar Orbiter \citep{Mueller2020so} also observed this event, providing HXR observations that reveal where these accelerated electrons deposit their energy in the chromosphere, the same atmospheric layer in which \ion{He}{I} D$_3$ forms.

The Fabry-Pérot system at IRSOL provides narrowband spectropolarimetric imaging by combining two voltage-tunable LiNbO$_3$ Fabry-Pérot etalons with a high-resolution Czerny-Turner spectrograph \citep{kleint2011fabryperot}. This configuration enables image acquisition with the option to scan across wavelengths. This setup has already been successfully applied in solar prominence observations with \ion{He}{I} D$_3$ by \citet{dicampli2020}. This instrumentation allows for quasi-monochromatic full-Stokes imaging with a spectral width of 35\,m\AA~across a FOV of approximately $65^{\prime\prime} \times 195^{\prime\prime}$ \citep{kleint2011fabryperot,dicampli2020}. For the 15 November 2024 M1 event observation, we prioritized obtaining images with minimal integration time, as well as avoiding scanning across many wavelengths. This approach was chosen due to the rapid dynamics of flares, as we aimed to avoid integrating together in the same image significant morphological changes of the flare. Consequently, during the 15 November 2024 M1 event, we configured the setup to observe one wavelength point in the \ion{He}{I} D$_3$ core (at 5876~\AA) and one in the nearby continuum (at 5879~\AA). The observing sequence was structured to repeat four \ion{He}{I} D$_3$ measurements followed by one measurement of the continuum throughout the duration of the flare. 
Unfortunately, the signal-to-noise ratio in linear polarization was too low to reveal any statistically significant flare-related temporal variations. We therefore restricted our analysis to Stokes \textit{I}.

\section{Data analysis}
\label{datan}
\subsection{3 May 2023 M7 event with slit-spectrograph}
\label{anaslit}
To estimate the position of the slit on the solar disk, we cross-compared the $\text{H}\alpha$ slit-jaw camera image taken during the observation with the one provided by the Udaipur observatory from the Global Oscillation Network Group \citep[GONG;][]{Harvey1996gong}. At the same time, we cross-compared the structures seen in the continuum and magnetogram data provided by HMI/SDO with the ZIMPOL full-Stokes slit-image. From this procedure, we obtained the helioprojective coordinates x=-620$^{\prime\prime}$, y=252$^{\prime\prime}$ for the center of the slit,
as shown in Fig.~\ref{fig:overview-figure}.

The raw data of the 3 May 2023 M7 event were reduced by applying the dark current, the polarization calibration, and the flat-field on Stokes \textit{I} only. For this observation, we employed the slow modulation technique, which suppresses systematic errors induced by the telescope as well as fringes. We note that the retarder foil employed for this measurement (a polycarbonate half-wave plate with 280 nm optical difference at 560 nm) provides a very high efficiency for Stokes \textit{Q} and \textit{U} corrections, but it has a low efficiency for Stokes \textit{V}. For this reason, we decided to not apply the slow-modulation corrections to Stokes $V$. Thus, in the reduced data there are uncorrected systematics, which are responsible for the nonzero value of the continuum in Stokes \textit{V}.
Figure~\ref{fig:spectral-profiles} illustrates the temporal evolution of the Stokes profiles, where we removed the polarized continuum in Stokes \textit{V} by subtracting its mean value.
The same procedure (continuum value subtraction) was used to determine the maximum of the circular polarization in the spectrum shown in the third panel of Fig.~\ref{fig:time-histories}.

For the analysis, we averaged along the slit the spectral profiles included in the spatial region associated with the flare, which we identified with the region of strong absorption profiles in \ion{He}{I} D$_3$. This allows us to significantly improve the signal-to-noise ratio of the linear polarization, up to a level of 5-6 $\sigma$. We averaged 12 pixels along the slit, corresponding to about 16.8 arcsec. Moreover, for the analysis of Stokes \textit{I}, we assumed each \ion{He}{I} D$_3$ profile in the time series to be the sum of a static and a dynamic profile, with the static one being the profile corresponding to the first time frame, which we considered to be in a pre-flare state. We then recovered the dynamic profiles by subtracting the static profile from the Stokes \textit{I} profile for each time frame. This method is particularly effective for the analysis of the flare of this event, because the subtraction removes the blending from the telluric lines and it highlights the difference between the pre-flare phase and the later phases. As an estimate of the Stokes \textit{I} continuum value, we chose the average of Stokes \textit{I} in the range of 5878.7--5879.0 ~\AA. The values of the absorption depth in \ion{He}{I} D$_3$ (see Fig.~\ref{fig:time-histories}, third panel) were computed from the Stokes \textit{I} pre-flare subtracted profiles as the difference between \ion{He}{I} D$_3$ line center and the continuum. For each time frame, we calculated the maximum degree of linear polarization, $P_{\max}$, to track its evolution during the solar flare (see Fig.~\ref{fig:time-histories}, third panel). This was done by applying a Savitzky-Golay filter to smooth the Stokes \textit{Q} and \textit{U} profiles, and then using the following formula:
\begin{equation}
    \label{poldeg}
    P_{\max} = \max_{\lambda} \left( \frac{\sqrt{Q(\lambda )^2 +U(\lambda)^2}}{I_c} \right)
.\end{equation}
This definition is not suitable for the case in which the line is in emission, because in that case $P_{\max}$ can exceed 100\%. This is not an issue for this study, since all the profiles we are considering exhibit absorption. 

To estimate the plasma bulk velocities associated with the \ion{He}{I} D$_3$ dynamic profiles (see Fig.~\ref{fig:time-histories}, bottom panel), we used the Hanle and Zeeman Light inversion code \citep[HaZeL;][]{ramos2008advanced}. HaZel allows one to accurately invert spectropolarimetric observations of \ion{He}{I} D$_3$ in optically thin prominences and filaments, retrieving information on both the magnetic field and bulk velocities. 
Since the scenario considered in this work is rather different, and not all the assumptions on which HaZeL is based on are strictly verified, we used it to estimate only the line-of-sight (LOS) bulk velocity, but not the magnetic field. 
Indeed, HaZel assumes an unpolarized, cylindrically symmetric pumping radiation field, which may not be accurate in the case of a flaring region, with nearby sunspots, as in the 3 May 2023 M7 event.
Moreover, HaZel does not account for radiative transfer effects within the considered emitting plasma structure (except along the considered LOS), as well as for the impact of collisions, which may have a significant effect in solar flares.

\subsection{15 November 2024 M1 event with Fabry-Pérot}
\label{sec:fabry-perot}

The 15 November 2024 M1 event was observed with the Fabry–Pérot setup, and the Stokes \textit{I} data were processed using dark-current correction, polarization calibration, and flat-field correction.
Since our goal was to compare the electron precipitation sites and the spatial distribution of the \ion{He}{I} D$_3$ profiles exhibiting strong absorption, we co-aligned SDO/AIA 1700~\AA, 304~\AA{}, and 94~\AA{} with STIX HXR images and the spatial regions of strong absorption in the \ion{He}{I} D$_3$ as observed at IRSOL with the Fabry-Pérot. More details on the SDO/AIA and STIX data reduction are given in the following subsection. For this comparison, we only considered the intensity image in the middle of the impulsive phase of the flare (12:14 UT), as previous studies have established a clear correlation between HXRs from the electron precipitation site and WL emission \citep[e.g.,][]{juan2012wl,sam2015white-light,kuhar2016wl,Battaglia2025gammaray} during the impulsive phase. 

\subsection{SDO/AIA, SDO/HMI, and Solar Orbiter/STIX}
\label{aiastix}
The data from the Atmospheric Imaging Assembly \citep[AIA;][]{2012SoPh..275...17L} and the Helioseismic and Magnetic Imager \citep[HMI;][]{2012SoPh..275..207S}, which are instruments aboard the Solar Dynamics Observatory \citep[SDO;][]{2012SoPh..275....3P}, were obtained from the Joint Science Operations Center (JSOC). The SDO/AIA maps were calibrated using the aiapy software \citep{2020JOSS....5.2801B}. The HMI continuum running difference shown in Fig.~\ref{fig:overview-figure} is part of a running difference sequence, whereby each image is generated by subtracting the previous image in time from the current one and then dividing by the previous image. According to \citet{kuhar2016wl}, the running sequence is mathematically defined as 
\begin{equation}
\label{rundif}
    S_{j+1} = \frac{I_{j+1} -I_j}{I_j},
\end{equation}
where $I_j$ is the SDO/HMI continuum image at time labeled $j$, with $j+1$ indicating the image acquired in the next time step. With the best time cadence, each publicly available SDO/HMI image is separated in time by 45 seconds.

The STIX observation data were downloaded from the STIX data center \citep{xiao2023datacenter}. 
For the reconstruction of the STIX image of the 15 November 2024 M1 event, we considered a time interval of one minute centered around the high-energy peak at 12:13:26 UT. The energy range only included nonthermal emission, from 16 to 28 keV.
The CLEAN algorithm \citep{1974A&AS...15..417H} was used, in which we applied natural weighting, and subcollimators from 3 to 10, with a beam width of 14.6 arcsec, corresponding to the resolution of subcollimators 3. 
Following the reconstruction, the STIX image was reprojected to the Earth perspective, enabling a direct comparison with the ZIMPOL Fabry-Pérot image as well as the SDO/AIA images. Image reprojection was performed using standard tools provided by SunPy \citep{2020ApJ...890...68S}. We then co-aligned the STIX image with the SDO/AIA 1700~\AA~map, by aligning the nonthermal sources with the flare ribbons observed in the UV map. 
To remove background contributions, the SDO/AIA 1700~\AA~image shown in Fig.~\ref{fig:overview-figure} was subtracted by another 1700~\AA~image before the start of the flare.

\begin{figure*}[htp!]
    \centering
    \vspace*{1cm}
    \includegraphics[width=0.99\linewidth]{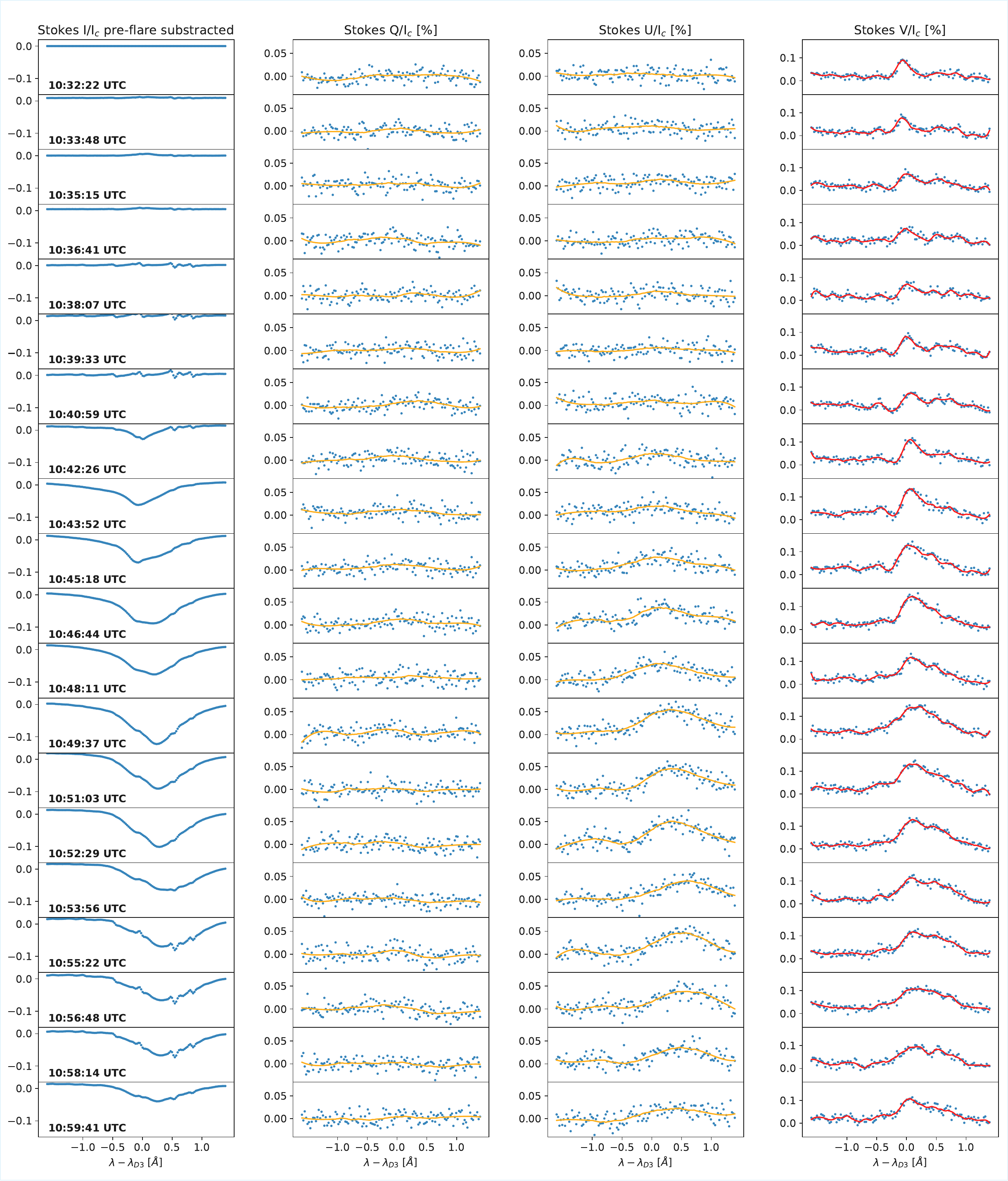}
    \caption{Time evolution of the Stokes signals for the 3 May 2023 M7 event.
    From left to right, we show the $I/I_c$ 
    (pre-flare subtracted, see Sect.~\ref{anaslit}), $Q/I_c$, 
    $U/I_c$, and $V/I_c$ 
    Stokes spectra averaged along the slit at each time frame of the observation, with $\lambda_{D_3}=5875.6~\AA$.
    We overplot the Savitzky-Golay smoothing of data points for $Q/I_c$ and $U/I_c$  
    (orange curves), and for $V/I_c$  
    (red curves).
    \bigskip}
    \label{fig:spectral-profiles}
\end{figure*}

\section{Results}
\label{sec:results}

\subsection{3 May 2023 M7 event with slit-spectrograph}
\label{res23}
From the M7 flare event on 3 May 2023, we obtained 20 time frames, which span about 30 minutes of measurements. These frames exhibit a rich evolution, which is shown in Fig.~\ref{fig:spectral-profiles}. Throughout the full observation, we do not observe any linear polarization in $Q/I_c$ above the noise level (check Sect. \ref{23overview} for the definition of positive Stokes $Q$).

\begin{figure}
    \centering
    \includegraphics[width=\linewidth]{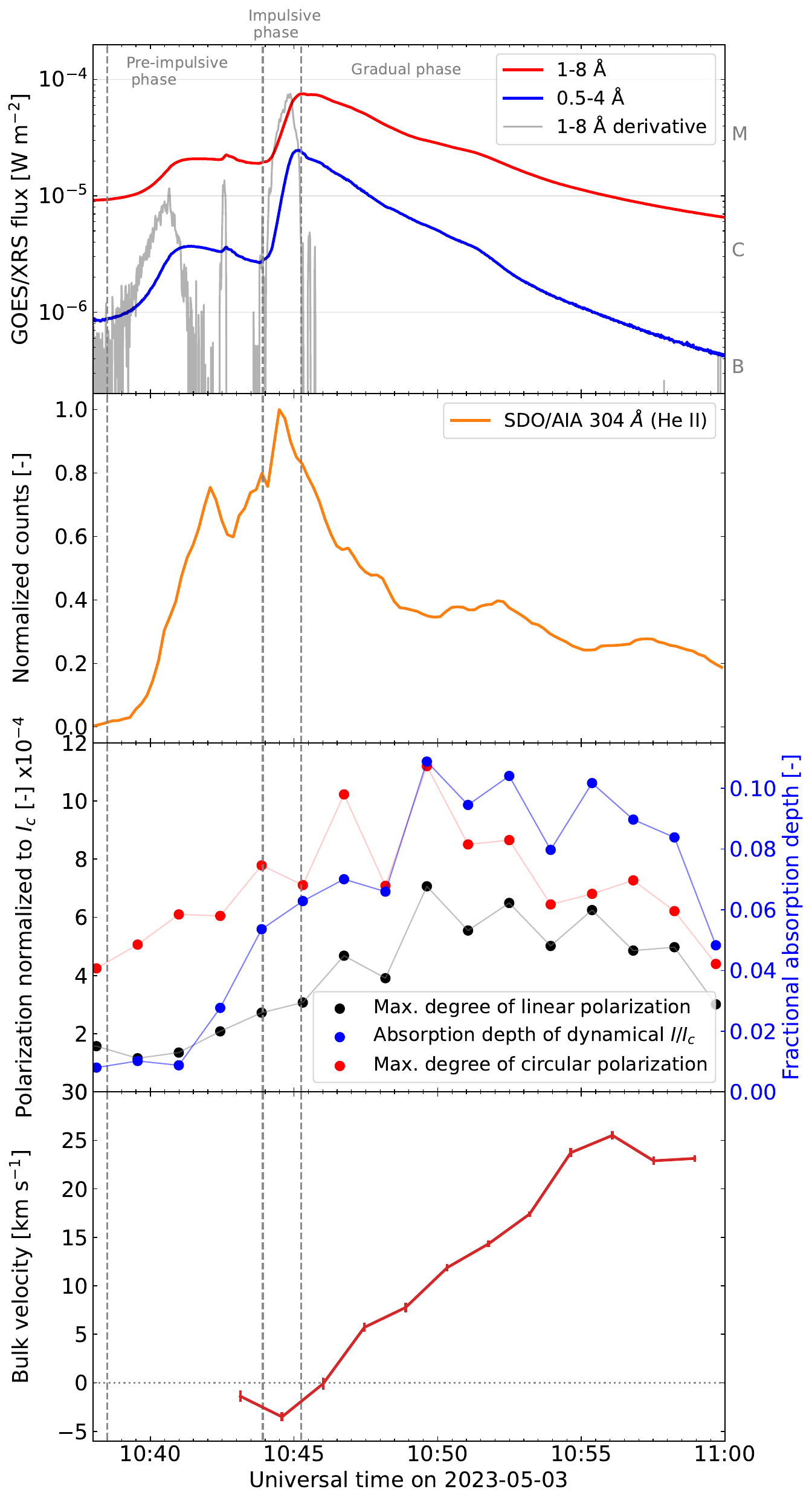}
    \caption{Time histories of 3 May 2023 M7 event. First panel (from the top): GOES X-rays flux. Second panel: Min-max normalized counts in SDO/AIA 304~\AA~channel. Third panel: Maximum in wavelength of the degree of linear polarization (black dots), absorption depth (blue dots), and maximum in wavelength of circular polarization (red dots) for \ion{He}{I} D$_3$. The value of the maximum of the degree of linear polarization was computed according to Eq.~\eqref{poldeg} and similarly the values of the maximum in circular polarization are a fraction of $I_c$. Fourth panel: LOS bulk velocities inferred by HaZel with uncertainties.}
    \label{fig:time-histories}
\end{figure}

In the first 10 minutes, we can see that the static part of the $I/I_c$ profiles dominates. In this regime, the \ion{He}{I} D$_3$ Stokes $V/I_c$ shows a single lobe shape, which cannot be explained by the Zeeman effect alone. During this static phase, Stokes $U/I_c$ does not show any significant signal above the noise level.
Approximately ten minutes after the measurements start, the dynamical phase sets in. The maximum in wavelength of the signal in Stokes $U/I_c$ and the absorption depth and the maximum in wavelength of Stokes $Q/I_c$ quickly increase, with the maximum in $U/I_c$ reaching a peak value of about $0.06 \%$. In the dynamical regime, the maximum of the value in Stokes $V/I_c$ is about twice the one seen in the linear polarization (Stokes $U/I_c$). Notably, the peak of the dynamical phase of \ion{He}{I} D$_3$ occurs about 5 minutes later than the peak in the GOES X-ray flux (see Fig.~\ref{fig:time-histories}). After the peak, the signal in $I/I_c$, $U/I_c$ and $V/I_c$ decreases, but at a slower rate compared to the raising phase.

\begin{figure}
    \centering
    \includegraphics[width=0.99\linewidth]{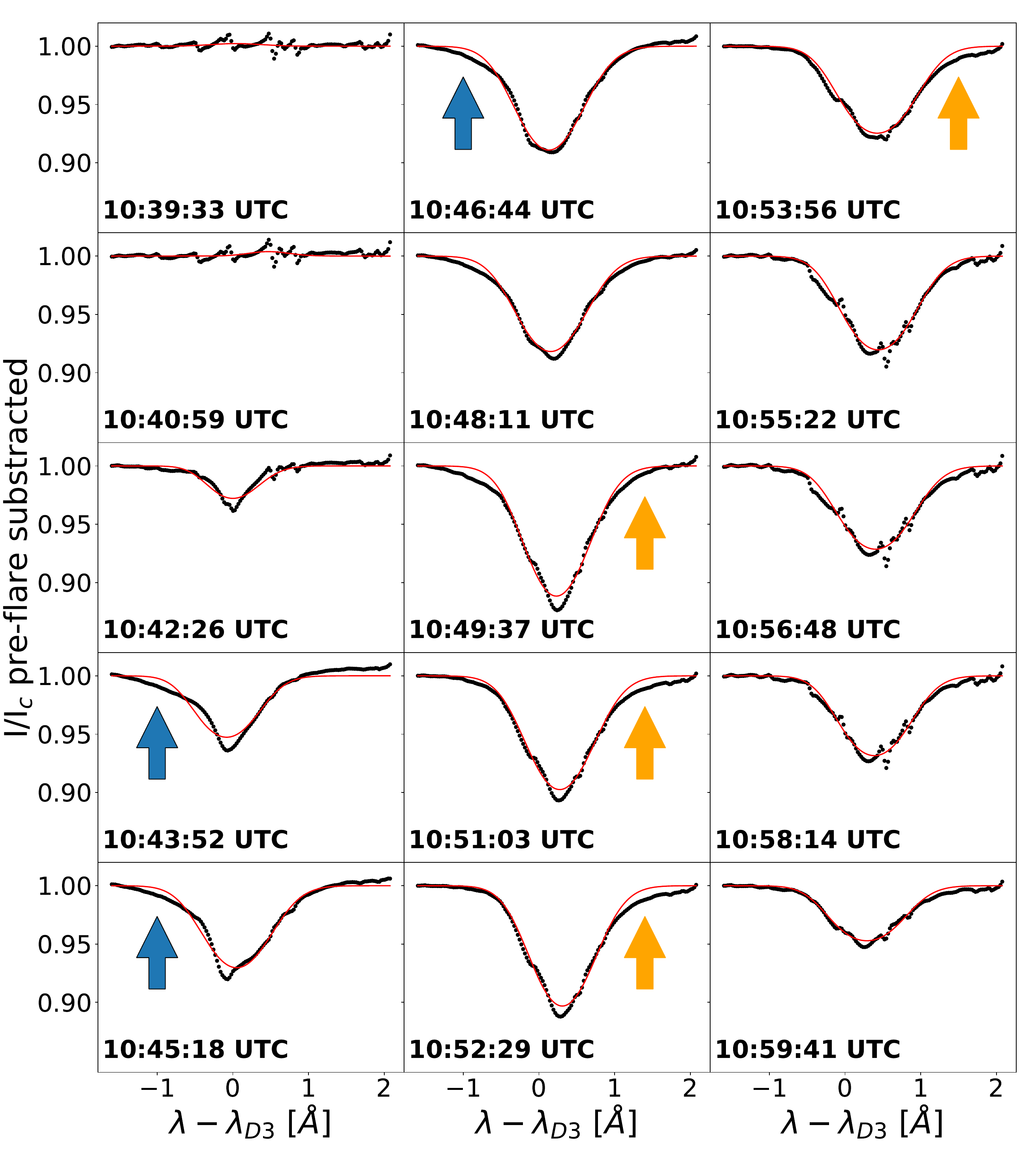}
    \caption{Time series of measured $I/I_c$ spectra (black dots) and the corresponding HaZeL fit (red lines) for the 3 May 2023 M7 event.
    Each image was processed by subtracting the chronologically first frame (10:33 UT, not shown here). 
    The blue and orange arrows indicate small discrepancies in the fit that might indicate the need to include in the modeling fainter components, blueshifted or redshifted, respectively, relative to the main component.}
    \label{fig:hazel-fit}
\end{figure}

The WL emission seen in the HMI continuum running differences lies outside the slit FOV (see Fig.~\ref{fig:overview-figure}). The SDO/AIA 304 trend closely follows the evolution of the X-ray flux as seen by GOES. This is particularly true in the impulsive phase of the event. To infer the LOS bulk velocities, we used HaZeL inversions of the observed \ion{He}{I} D$_3$ pre-flare subtracted Stokes $I/I_c$ profiles. For the inversion, we employed a single slab-model (one velocity component), which means that significant discrepancies between the data and the fit Gaussian profiles are indications of additional velocity components in the \ion{He}{I} D$_3$ line or indications of physical effects that are not taken into account in the theoretical model used by HaZeL.
Figure~\ref{fig:hazel-fit} shows the time series of the measured $I/I_c$ spectra and the corresponding HaZeL fit, which agree reasonably well. Since the bulk velocities are inferred from the spectral location of the line core, employing a more refined fitting procedure would not be expected to produce substantial changes in the derived velocities.

Figure~\ref{fig:time-histories} illustrates the retrieved LOS bulk velocities, which show a weak upward plasma displacement of about -5 km/s at the time of the peak of the flare X-ray flux. After the GOES peak, the plasma LOS velocity grows in the opposite direction (downward motion). About ten minutes after the peak, it reaches the maximum, which is about +25 km/s, and then it starts to slowly decay. From the analysis of the Gaussian fitting in the inversion process, we can identify the presence of discrepancies between the data and the fit profiles, which appear at 10:44 UT spectrally blueshifted with respect to the core of the main component, but become redshifted from 10:49 UT (see Fig.~\ref{fig:hazel-fit}). Although the quality of the fit shown in Fig. \ref{fig:hazel-fit} could be improved using two slabs within HaZeL, the introduction of additional slabs would increase the geometric complexity and the number of free parameters without providing key information for our temporal analysis. Our primary goal is to recover the temporal evolution of the mean bulk velocities and for this reason we believe that a single-slab inversion is enough in this context. 

\subsection{15 November 2024 M1 event with Fabry-Pérot}

\begin{figure*}
    \centering
    \includegraphics[width=\linewidth]{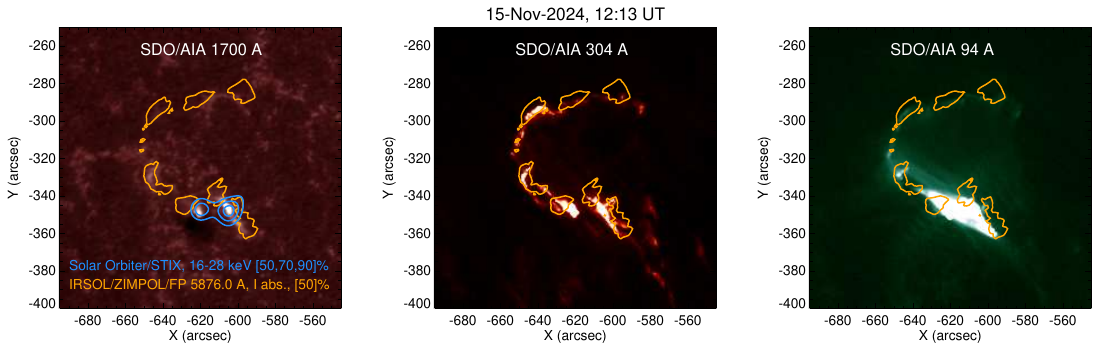}
    \caption{Comparison of SDO/AIA maps with STIX and ZIMPOL Fabry-Pérot observations. The orange contours represent the 50\% of the maximum Stokes \textit{I} in the ZIMPOL observations of the 15 November 2024 M1 event during the impulsive phase using the Fabry-Pérot setup.
    For context, from left to right we show the SDO/AIA 1700~\AA~(WL emission), SDO/AIA 304~\AA~(flare ribbons), and SDO/AIA 94~\AA~(flare loop) maps. The blue contours represent the STIX HXR images, indicating the electron precipitation sites, which are co-spatial with the WL emission in the SDO/AIA 1700~\AA~map.}
    \label{fig:Fabry-Pérot}
\end{figure*}

For the observation of the 15 November 2024 M1 event, we compared the Fabry-Pérot Stokes \textit{I} image with SDO/AIA 1700~\AA, SDO/AIA 304~\AA~SDO/AIA 94~\AA, and STIX images (see Fig.~\ref{fig:Fabry-Pérot}). We can observe that the stronger \ion{He}{I} D$_3$ absorption profiles coincide spatially with the location of the flare ribbons.

Interestingly, the strong absorption in Stokes \textit{I} does not align with the location of the HXRs, which indicates the location of the electron precipitation site. A possible explanation is that the line may have transitioned into emission, thereby compensating for the absorption. Previous studies have indeed suggested that this line goes into emission at the electron precipitation site \citep[e.g.,][]{Liu2013HeID3flare,libbrecht2019chromospheric}. Unfortunately, since with the present setup we observed only a single wavelength point in the line, we cannot verify this spectrally.

\section{Discussions}
\label{sec:discussions}
\subsection{Temporal evolution and formation mechanisms}
\label{disabs}

As is outlined in Sect.~\ref{res23},
during the 3 May 2023 M7 event, the temporal evolution of intensity in the SDO/AIA 304~\AA~channel strongly correlates with the GOES X-ray flux, suggesting that the energy released by the solar flare is ionizing the helium atoms, and hence increasing the density of ionized helium. In the decay phase of the X-ray flux, when the plasma temperatures decrease and the nonthermal processes have ended, we instead have helium recombination. Indeed, the quick raising in absorption depth of the \ion{He}{I} D$_3$ line, which indicates formation of ortho-helium, is temporally correlated with the decay in SDO/AIA 304~\AA.
For this event, it is difficult to identify whether the ionization-recombination is driven by collisions or UV illumination. In the simulation presented in \citet{kerr2021he},
nonthermal collisional ionization is identified as the primary mechanism in the early phase of the solar flare. However, as was already discussed in Sect.~\ref{res23}, our slit FOV is away from the location of WL emission in the HMI continuum running differences (see Fig.~\ref{fig:overview-figure}), which may indicate that the electron flux impacting the chromosphere at the slit FOV location is not enough to produce WL emission. This suggests that the dominant heating mechanism here may differ. Additionally, it is also possible that plasma displacement moved into the slit FOV the ortho-helium previously formed through collisional ionization–recombination in the region of high-flux electron precipitation.

We observe high correlation in time between the absorption depth of \ion{He}{I} D$_3$, its maximum in degree of linear polarization, and its maximum in circular polarization. This suggests that the evolution in ortho-helium densities is the main driver of the observed changes in time of the \ion{He}{I} D$_3$ Stokes profiles. In particular, the sharp increase is caused by the sudden energy release from the flare event, while the very long ortho-helium lifetime of about 2h \citep[e.g.,][]{hodgman2009metastable} explains the slow decay observed in the time histories of the Stokes profiles (see Fig.~\ref{fig:spectral-profiles}). However, to explain the variations in the Stokes profiles we cannot exclude contributions from the temporal evolution of the atmospheric thermal stratification and contributions from bulk velocity gradients.

The Fabry-Pérot observation of the 15 November 2024 M1 event shows that the spatial region of the strongest absorption in the \ion{He}{I} D$_3$ line correlates well with the flare ribbons as seen by SDO/AIA 304 \AA~(middle panel of Fig.~\ref{fig:Fabry-Pérot}). Thus, we associated the region traced by strong absorption in \ion{He}{I} D$_3$ with the ribbons of the flare. We note that this does not rule out possible absorption originating from the flare loop \citep[as reported by][]{libbrecht2019chromospheric}, since the stronger absorption in the ribbons may mask weaker signals, as was previously reported in \citet{Liu2013HeID3flare}.

\subsection{Polarization signals in \ion{He}{I} D$_3$}
\label{poldis}

The statistical study by \citet{kuhar2016wl} shows that there exists a high correlation between the location of HXRs and the location of WL emission as seen by SDO/HMI. In particular, the correlation is higher with X-rays produced by very energetic electrons (>50 KeV). In our observation, the WL observed in the HMI continuum running differences does not spatially correlate with the slit FOV. Moreover, we do not find a correlation in time between the peak of the impulsive phase and the peak in the maximum of \ion{He}{I} D$_3$ linear polarization. Therefore, the location and especially the timing of the observation suggest that the large-scale anisotropic collisions of flare-accelerated particles (i.e., impact polarization) may not be the dominant mechanism generating the observed linear polarization of the 3 May 2023 M7 event. The observation performed in this particular slit FOV does not exclude the presence of impact polarization in \ion{He}{I} D$_3$ in another region or that polarization is produced and destroyed by local anisotropic collisions caused by small-scale events.

We can exclude significant contributions from the Zeeman effect on the linear polarization because we observe that the polarization signal has one single lobe instead of three. The Zeeman effect can in principle produce a single lobe signal in the linear polarization if there are multiple velocity components. However, to maintain a single-lobe shape of the signal, it is required that the bulk velocities remain constant throughout our observation. This is unlikely considering that the bulk velocities derived through HaZeL show significant changes in time (see Fig.~\ref{fig:time-histories}, bottom panel). Moreover, the Zeeman effect in ARs usually produces a stronger signal in Stokes $V$ than in the linear polarization. However, this does not occur in our observation. Namely, in the initial time frames, the maximum in Stokes $V$ signal is about twice the one in linear polarization. This ratio becomes smaller in the time frames from 10:53 UT to the end of the measurements. Thus, we argue that the mechanism behind the linear polarization is probably scattering polarization due to anisotropy in the pumping radiation field. In our observation, the anisotropy in the radiation field could be produced by the vertical stratification of the solar atmosphere, the presence of sunspots in the underlying photosphere, or the gradient in illumination at the edges of the solar flare ribbons \citep{vstvepan2013scattering}. We note that the first two sources do not change appreciably during the observation and therefore would not explain the observed time variation in the Stokes profiles.
On the other hand, this time variation seems to be better explained by variations in the density of ortho-helium (see Sect.~\ref{disabs}) rather than in the anisotropy. In this sense, it is difficult to identify which one of three possible sources of anisotropy is the most important in our observation.

The Zeeman effect alone produces perfectly antisymmetric Stokes $V$ profiles, and thus zero net circular polarization (NCP). Stokes $V$ profiles with NCP in He~{\sc i} D$_3$ have already been observed in flares \citep[e.g.,][]{lozitska2024unique, libbrecht2019chromospheric}. In the analysis of the event presented in \citet{libbrecht2019chromospheric}, it is concluded that the NCP can be explained solely by the presence of an additional velocity component (in the presence of a magnetic field). 

In general, there are two main ways to generate asymmetries in Stokes $V$ and, in particular, single-lobe profiles such as the ones we observed in the 3 May 2023 M7 event. The first one is the presence of correlations between velocity and magnetic field gradients \citep[e.g.,][]{1975A&A....41..183I}. Notably, the single-lobe Stokes $V$ profiles resulting from this scenario should have a width that is approximately half that of Stokes $I$ \citep[e.g.,][]{gonzalez2012anomalous}. In our observation, this is not the case, especially for the profiles in the time frames > 10 minutes from the start of the observation (see Fig.~\ref{fig:spectral-profiles}). 
The second way to generate asymmetries in Stokes $V$ is through the presence of atomic orientation; namely, population imbalances and coherence between Zeeman sublevels with magnetic quantum numbers $+M$ and $-M$. This second scenario moves the problem to the identification of possible mechanisms through which atomic orientation can be created. For this, there are two main mechanisms: 1) the alignment-to-orientation (A-O) conversion mechanism \citep[e.g.,][]{landi2004polarization}, and 2) a differential illumination of the $\sigma^\pm$ components (i.e., the components with $\Delta M = \pm 1$) of the considered line. The A-O mechanism is particularly important in \ion{He}{I} D$_3$, as it starts operating for magnetic field strengths above 10\,G and has maximum efficacy around 100\,G \citep[e.g.,][]{casini2008astrophysical}. In our observation, we note that HaZeL inversions of the observation estimate the magnetic field to be around 50 G. However, this value seems way too low considering that the observation occurred above a sunspot group. It is possible that the HaZeL code can fit (with one slab) significant NCP only by invoking the A-O conversion mechanism, and hence deriving a magnetic field value that is around 100 G.

Concerning the differential illumination, there are two main scenarios \citep[e.g.,][]{gonzalez2012anomalous}: (a) pumping the atomic system with circularly polarized radiation, and (b) splitting the Zeeman components of the line (e.g., via a magnetic field) and pumping them with different radiation fields. The first scenario can be expected when the underlying photosphere is highly magnetized. This is indeed the case of our observation, where the slit was placed very close to a group of sunspots. On the other hand, we note that the sunspots have opposite polarities, and this could reduce the efficacy of this mechanism due to cancellation effects. Scenario (b) can take place when a magnetic field is present and the atoms are moving so that their absorption profiles are Doppler shifted with respect to the pumping radiation field (assumed symmetric with respect to the unshifted line-center frequency). This scenario cannot be ruled out either, as our inversions with the HaZel code indicate significant bulk velocities (up to about 20\,km/s) and magnetic fields in the photosphere (SDO/HMI magnetogram) in the range of 500-1000 G. From the information available, it is not trivial to better understand which of the aforementioned mechanisms contributes the most to the observed signal. Indeed, this would require a more quantitative modeling that would go beyond the scope of this paper. 

\subsection{LOS plasma bulk velocities}
\label{veldis}
In the observation of the 3 May 2023 M7 event, we found strong downflows along the LOS in the decaying phase of the flare with at least two velocity components. The most prominent component has values around 20 km/s (from 10:55 UT). We also observe an upflow of about 5 km/s in the impulsive phase of the flare, which seems to be on the lower edge of the typical gentle evaporation velocities observed during the impulsive phase of flares \citep[e.g.,][]{Milligan2006gentleevaporation}. However, the fact that the flare has been observed near the limb contributes to lower this estimate, because the LOS is tilted with respect to the local vertical. On the other hand, gentle evaporation is associated with flares with low flux of nonthermal electrons. This applies for our observation, because from the inspection of the time profiles provided by the Hard X-ray Imager \citep[HXI;][]{Zhang2019hxi} aboard the Advanced Space-Based Solar Observatory \citep[ASOS;][]{Gan2019asos}, there is indeed no detectable signal above 30 keV, and for an M GOES class flare this suggests a low flux of nonthermal electrons.

\section{Conclusions and outlook}
\label{sec:conclusions}
Using the ZIMPOL polarimeter at IRSOL observatory in Locarno (Switzerland), we collected \ion{He}{I} D$_3$
spectropolarimetric observations during an M7 GOES class flare on 3 May 2023 and intensity images during an M1 GOES class flare on 15 November 2024. From the analysis of the measurements, we conclude that:

\begin{enumerate}
    \item ZIMPOL is nicely suited to study polarization in flares down to a precision level of $10^{-4}$ in fractional polarization that is hardly accessible with other instruments.
    This capability is enabled by the recent implementation at IRSOL of combined fast and slow modulation techniques \citep{zeuner2022enhancing}, which effectively suppress polarization artifacts.
    \item For the first time, we unequivocally detected linear polarization in \ion{He}{I} D$_3$ during a flare with peak amplitudes of $6\cdot10^{-4}$ in $U/I_c$, while also tracking the temporal evolution of the polarization signal.
    \item The \ion{He}{I} D$_3$ linear polarization signal we observed is likely not dominated by impact polarization, as it shows neither spatial correspondence with the electron precipitation sites nor temporal synchronization with the impulsive phase. However, small-scale events may still drive anisotropic collisions locally, meaning impact polarization could still provide some contribution.
    \item The detected \ion{He}{I} D$_3$ linear polarization signals are likely produced by scattering processes in the presence of an anisotropic radiation field. We can exclude a relevant contribution from the Zeeman effect.
    \item The time variation in Stokes $I/I_c$, $U/I_c$, and $V/I_c$ could be explained by the variation in the density of ortho-helium, which has been produced by the solar flare through ionization recombination. However, we cannot rule out that other mechanisms, such as the temporal evolution of the atmospheric thermal stratification and of plasma bulk velocities, could also affect the observed signals.
    \item There is a correspondence between the ZIMPOL observation with the Fabry-Pérot system (15 November 2024) showing absorption in \ion{He}{I} D$_3$ and the emission observed in the SDO/AIA 304~\AA~map during the impulsive phase. We therefore conclude that the strongest absorption signals in \ion{He}{I} D$_3$ originate from the ribbons of the flares.
\end{enumerate}

Ultimately, spectropolarimetry in flares can provide valuable constraints on flare models and may also offer insights into particle acceleration mechanisms. However, to fully exploit this diagnostics, one has to capture the impulsive phase at the appropriate locations; namely, at the precipitation sites of electrons, protons, and ions.

We finally note that this study stems from an ongoing wider project on spectropolarimetric observations of solar flares with ZIMPOL.
In this regard, the IRSOL observatory is currently undergoing upgrades to improve the likelihood of capturing the impulsive phase of flares and to facilitate accurate slit placement. In parallel, the regular deployment of ZIMPOL at the 1.5-m GREGOR telescope \citep[e.g.,][]{Bianda2018gregor, Dhara2019zimpolgregor} will enable flare observations with both a high polarimetric sensitivity and a high spatial resolution. This capability is crucial for investigating the polarization patterns within fine ribbon structures, such as the newly discovered ribbon fronts \citep[e.g.,][]{Liu2018ribbonfronts, Polito2023ribbonfronts, Kerr2024ribbonfronts}, thereby helping to constrain flare models.

\begin{acknowledgements}
    We thank Dr. F. Zeuner and Dr. M. Bianda for the valuable discussions and technical assistance with the IRSOL observatory instrumentation. We acknowledge and thank the anonymous referee for the valuable inputs and encouraging comments.
    
    IRSOL is supported by the Swiss Confederation (SERI), Canton Ticino, the city of Locarno and the local municipalities. Solar Orbiter is a space mission of international collaboration between ESA and NASA, operated by ESA. The STIX instrument is an international collaboration between Switzerland, Poland, France, Czech Republic, Germany, Austria, Ireland, and Italy. FV, AFB, and RR are supported by the Swiss National Science Foundation (SNSF) grant 200020\_213147. 
    LB and GJ acknowledge support from SNSF through grant 231308.
    J.\v{S}. acknowledges the financial support from project \mbox{RVO:67985815} of the Astronomical Institute of the Czech Academy of Sciences.
\end{acknowledgements}

\bibliographystyle{aa}
\bibliography{biblio}

\end{document}